\documentclass{aa}
\usepackage{psfig}

\def\Ni{\noindent}
\def\ros{{\sl ROSAT}}
\def\etal{{et\,al.}}

\def\ergcm{\hbox{erg cm$^{-2}$ s$^{-1}$ }}

\def\grad{$^\circ$}

\def\degs{\ifmmode ^{\circ}\else$^{\circ}$\fi}
\def\amin{\ifmmode ^{\prime}\else$^{\prime}$\fi}
\def\asec{\ifmmode ^{\prime\prime}\else$^{\prime\prime}$\fi}
\def\fss{\hbox{$.\!\!^{\rm s}$}}        
\def\h{$^{\rm h}$}
\def\m{$^{\rm m}$}

\def\rxj{RX\,J1016.9--4103}
\def\59{RX\,J1016}

\begin{document}
 
   \thesaurus{06         
              (02.01.2;  
               08.09.2;  
               08.13.1;  
               08.14.2;  
               13.25.5)} 

 \title{\rxj: A new soft X-ray polar in the period gap\thanks{Partly based on 
       observations collected during MPI time at the 2.2 m telescope of the 
       European Southern Observatory, La Silla, Chile.}}

   \author{Jochen Greiner  \and Robert Schwarz}

   \offprints{Greiner, jgreiner@aip.de}
 
  \institute{Astrophysical Institute
          Potsdam, An der Sternwarte 16, 14482 Potsdam, Germany}

   \date{Received 18 August 1998; accepted ?? September 1998}
 
   \maketitle

   \markboth{J. Greiner \& R. Schwarz: The new polar \rxj}{J. Greiner 
      \& R. Schwarz:The new polar \rxj}

   \begin{abstract}
We have discovered a new AM Her system as the optical counterpart of the
\ros\ All-Sky-Survey source \rxj\ ($\equiv$ 1RXS J101659.4-410332). The X-ray 
spectrum is very soft and the X-ray intensity is strongly modulated with the
orbital period.
Optical photometric and spectroscopic follow-up observations reveal
a synchronously rotating binary with an orbital period of 134 min, placing
\rxj\ in the period gap. The strength of the TiO bands suggests a secondary
spectral type  later than M3 V and a distance of 615$\pm$150 pc.
Based on two clearly visible broad humps in the optical spectrum 
(interpreted as cyclotron features) a magnetic field strength of 52 MG is
deduced thus proving the polar classification.

      \keywords{X-rays: stars -- stars: cataclysmic variables -- 
           accretion, accretion discs -- stars: magnetic fields  -- 
           stars: individual: \rxj\ $\equiv$ 1RXS J101659.4-410332}

   \end{abstract}
 
\section{Introduction}

 \begin{table}[th]
 \caption{Log of observations}
 \vspace{-0.2cm}
 \begin{tabular}{lllr}
      \hline
      \noalign{\smallskip}
 Telescope$^{a}$  & ~~~Date  &  Energy Band &  $T_{\rm int}$ \\
 &&  &  (sec) \\
      \noalign{\smallskip}
      \hline
      \noalign{\smallskip}
 \multicolumn{4}{c}{\bf X-ray} \\
 ROSAT P & 1990 Dec 09--11    & 0.1--2.4 keV  & 430   \\
 ROSAT P & 1993 Nov 30--Dec 4 & 0.1--2.4 keV & 6760  \\
 ROSAT P & 1994 Jun 01        & 0.1--2.4 keV & 2930  \\
 ROSAT H & 1997 Dec 08        & 0.1--2.4 keV & 1800  \\
   \noalign{\smallskip}
 \multicolumn{4}{c}{\bf optical spectroscopy} \\
 ESO 2.2 m & 1995 Mar 26   & 3000--9000 \AA & 1800 \\
   \noalign{\smallskip}
 \multicolumn{4}{c}{\bf optical photometry} \\
 ESO 2.2 m & 1995 Mar 25--27   & B, R  & 600, 180$\!\!$ \\
 \noalign{\smallskip}
 \hline
 \noalign{\smallskip}
 \end{tabular}

\noindent{\small $^{a}$ The abbreviations have the following meanings:
  ROSAT P or H = PSPC or HRI detectors onboard the ROSAT satellite,
  ESO = European Southern Observatory, La Silla/Chile.}
   \label{log}
 \end{table}

AM Her type variables  are a subgroup of cataclysmic variables (CVs)
in which the magnetic field of the white dwarf controls the geometry 
of the material flow between the main-sequence donor and the white dwarf
primary as well as synchronizes the white dwarf spin period
with the binary orbital period (see e.g. Warner 1995 for a detailed review). 
The inflow of matter along the magnetic field lines (of one or
occasionally also two magnetic poles) is decelerated above the white dwarf
surface producing a shock front. This region is thought to emit hard X-rays
(usually modelled in terms of thermal bremsstrahlung of 10--20 keV)
and polarized cyclotron radiation (hence these systems are named also polars) 
in the IR to UV range. 
In addition, a strong soft component has been frequently observed from polars
which is thought to arise from the heated accretion pole
(usually modelled in terms of blackbody emission with 20--50 eV temperature).

It is this soft X-ray component which has led to the discovery of a few
dozen new polars by \ros\ observations over the last 8 years, 
most notably the \ros\ all-sky survey (e.g. Beuermann \& Burwitz 1995 
for a first summary). Based on these discoveries, 
the strength of the soft component (or more precisely the ratio of the
blackbody and the bremsstrahlung component) has been found to increase with
the magnetic field strength of the white dwarf. 
Since simple reprocession of the hard
component cannot produce the observed soft X-ray fluxes,
a scenario of high-$\dot{m}$, blobby accretion has been proposed as a 
mechanism to explain the soft excess (Kujpers \& Pringle 1982).
In a different approach, bremsstrahlung emission is gradually replaced
by more efficient cyclotron cooling at high magnetic field strengths
(Woelk \& Beuermann 1996).

The source described here has been
discovered as a result of a systematic survey for supersoft X-ray sources in
the all-sky survey data (see Greiner 1996 for details of this survey)
which revealed a large number of CVs and single white dwarfs. 
Other polars identified from this sample include
V844 Her = RX\,J1802.1+1804 (Greiner, Remillard and Motch 1995, 1998),
RS Cae = RX\,J0453.4--4213 (Burwitz \etal\ 1996), and
RX\,J1724.0+4114 (Greiner, Schwarz and Wenzel 1998).
In this paper we present photometric, spectroscopic
and X-ray observations (summarized in Tab.~\ref{log})
which led to the discovery of another polar, \rxj\ 
(henceforth referred to as \59).

  \begin{figure}
       \vbox{\psfig{figure=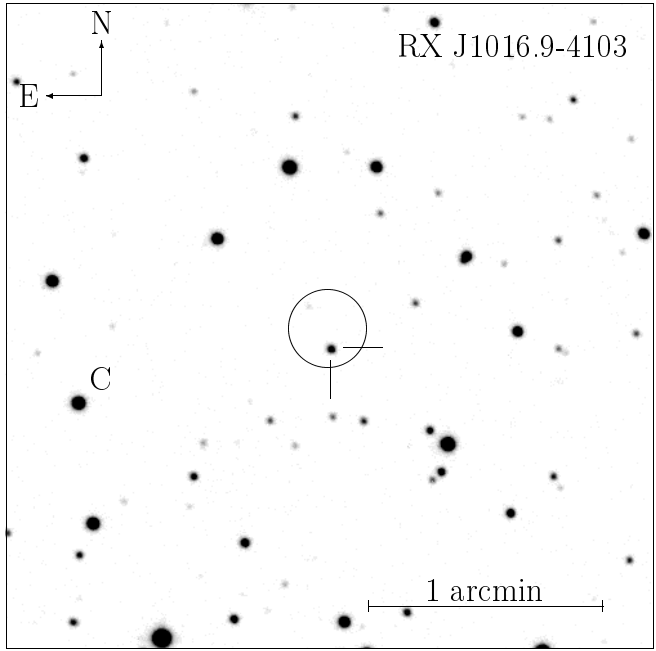,width=8.7cm}}\par
      \caption[fc]{A 10 min $B$-band CCD image of \rxj\ obtained with the
        2.2 m telescope at ESO. The circle denotes the 3$\sigma$ X-ray 
        error circle derived from the HRI pointing (10\asec\ radius). 
        The optical position is measured as: R.A. (2000.0) = 10\h16\m58\fss9,
        Decl. (2000.0) = --41\grad 03\amin 45\asec\ ($\pm$1\asec).
        The bright star at the left denoted with ``C'' has been used
        as comparison star for the photometry.
          }
      \label{fc}
\end{figure}

\section{X-ray discovery and identification}

\59\ was scanned during the \ros\ all-sky-survey over a period of 4 days
in December 1990 for a total observing time of 430 sec. 
Its mean count rate in the \ros\
position-sensitive proportional counter (PSPC) was 0.13 cts/s,
and the hardness ratio $H\!R1=-0.95\pm0.12$ where $H\!R1$ is defined as 
(H--S)/(H+S), with H (S) being the counts above (below) 0.4 keV over
the full PSPC range of 0.1--2.4 keV.

\begin{figure}  [t]
  \psfig{figure=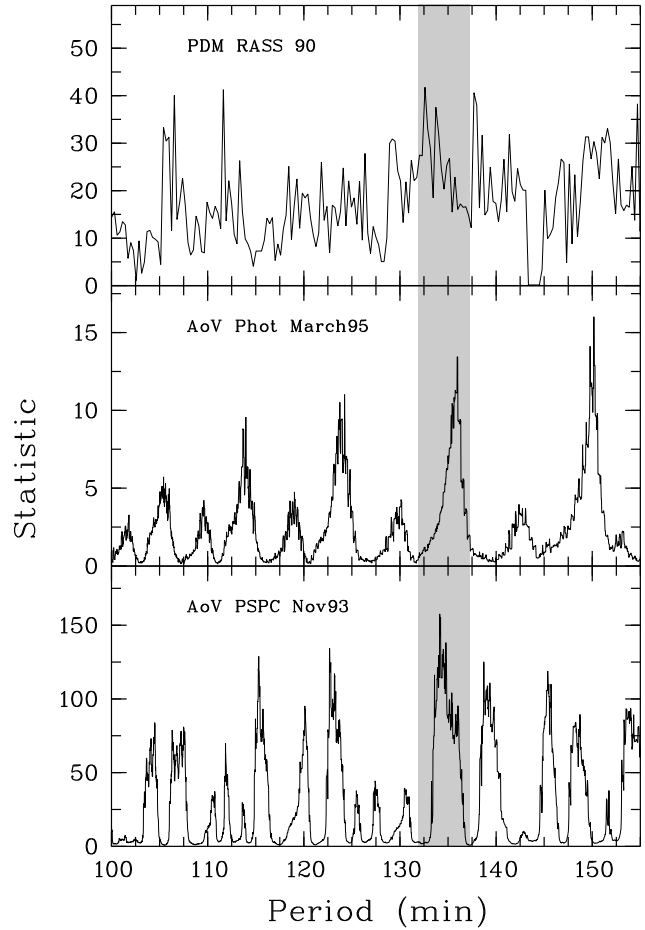,width=0.98\columnwidth,%
      bbllx=35pt,bblly=45pt,bburx=530pt,bbury=770pt,clip=}
  \caption[perio]{
     Results of the period-search using the analysis-of-variance method.
     The adopted period of 134$\pm$3 min is marked as the shaded region.}
    \label{perio}
\end{figure}

\begin{figure}
    \psfig{figure=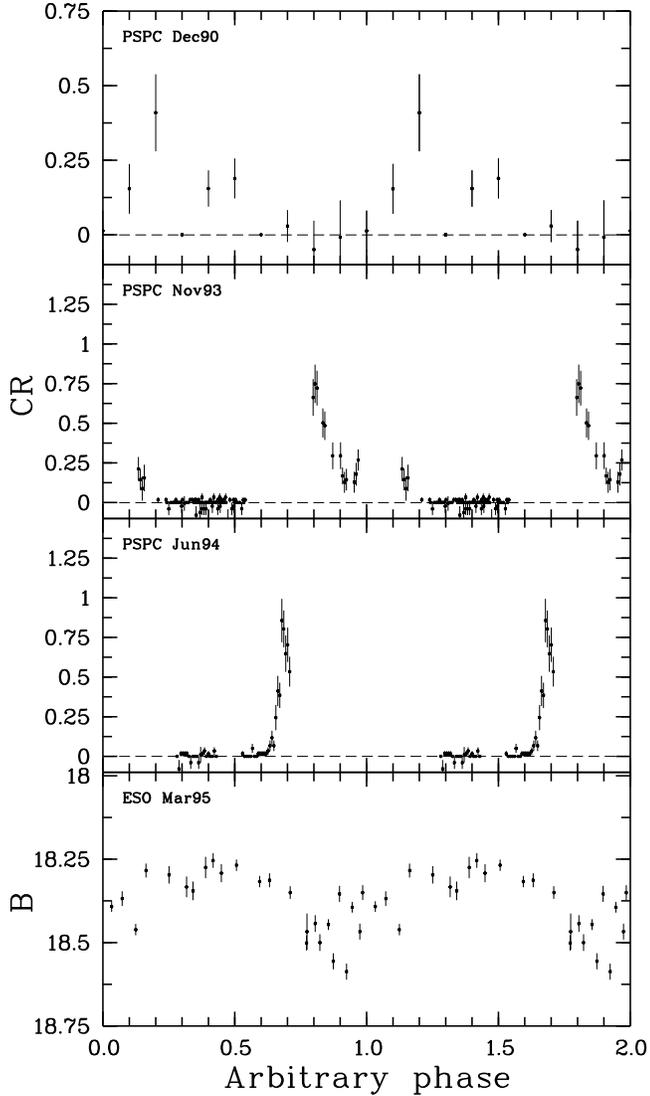,width=0.98\columnwidth,%
      bbllx=50pt,bblly=90pt,bburx=440pt,bbury=760pt,clip=}
   \caption[olc]{
   Optical and X-ray light curves of \59\ plotted over orbital phase for
   a period of $P=134.4$ min.
   Note that the phase designation of the middle two panels is arbitrary
   with respect to the all-sky survey (top panel) and optical (bottom panel)
   data. Data are plotted twice for clarity and units are PSPC cts/s  
   for the upper three panels and $B$  magnitudes for the bottom panel.}
  \label{lc}
\end{figure}

For the timing analysis the source photons were extracted with a radius of
4\amin. The background was chosen at the same ecliptic longitude at
$\approx$1\grad~ distance, corresponding to background photons collected
typically 15 sec before or after the time of the source photons. Standard
corrections were applied using the dedicated EXSAS software package
(Zimmermann \etal\ 1994). The RASS light curve folded
over the best-fit period as derived in section 3 is shown in
the upper panel of Fig. \ref{lc}.
The X-ray flux shows  100\% modulation with a peak count rate of $\sim$0.4
cts/s and a pronounced long faint-phase where the X-ray flux is nearly
zero (formal count rate of $0.05\pm 0.05$ cts/s).

The soft X-ray spectrum and strong variability immediately suggested 
the cataclysmic variable nature. Indeed, using the best-fit X-ray coordinate
derived from the all-sky survey data only one optical object brighter than 
20$^{\rm m}$ was found within the error circle (see Fig. \ref{fc}) which
later was identified spectroscopically as a polar (see section 4).

Dedicated follow-up pointed \ros\ observations were performed in November 
30 -- December 4, 1993 and June 1, 1994 with the PSPC and on December 8, 1997 
with the high-resolution imager (HRI). Table \ref{xlog}
summarizes the relevant numbers of these measurements including the
total number of detected counts (column 3), the mean count rate (column 4),
the hardness ratio (column 5), the coverage of the orbital phases (column 6),
the off-axis angle (column 7) and the best-fit X-ray positions (column 8).

\begin{table*}
\caption{Basic numbers of the X-ray observations of \rxj}
\begin{tabular}{lccccccc}
\hline \noalign{\smallskip}
      ~~~~~~Date & Obs-ID$^{(1)}$ & N$_{\rm cts}$ & mean count & $HR1$ & phase 
            & offaxis & X-ray position \\
                 &                &              & rate (cts/s) &   & coverage
            & angle  & (equinox 2000.0) \\
\noalign{\smallskip} \hline \noalign{\smallskip}
    1990 Dec 09--12$^{(1)}$ & -- & ~\,57  & 0.13     
         & --0.95$\pm$0.12 & ~~\,5\% & 0\amin--52\amin 
         & 10\h16\m58\fss2\,--41\grad03\amin38\asec\ $\pm$ 15\asec \\
    1993 Nov 30--Dec 4 & 201605p & 430 & 0.06
         & --0.98$\pm$0.01 & ~\,55\% & 0\farcm50
         & 10\h16\m59\fss8\,--41\grad03\amin52\asec\ $\pm$ 25\asec \\
    1994 Jun 1  & 201605p-1 & 352 & 0.12
         & --0.98$\pm$0.01 & ~\,30\% & 0\farcm32 
         & 10\h16\m59\fss2\,--41\grad03\amin44\asec\ $\pm$ 25\asec \\
    1997 Dec 08 & 300580h & ~~\,3 & ~~~0.0019$^{(3)}$
         &  --             & ~\,20\% & 0\farcm21
         & 10\h16\m58\fss9\,--41\grad03\amin40\asec\ $\pm$ 10\asec \\
      \hline 
      \noalign{\smallskip} 
      \end{tabular}
   \noindent{\Ni\small $^{(1)}$ The letters after the number denote the
                                 detector: h = HRI, p = PSPC.\\ 
                       $^{(2)}$ For the position determination only photons
                                in the energy range 0.25--0.5 keV have been 
                                used to avoid position degradation due to
                                ghost images. \\
                       $^{(3)}$ Note  the lower sensitivity of the
                                HRI at soft energies by a factor of 7.8 as
                                compared to the PSPC.
            }
    \label{xlog}
   \end{table*}

\section{Optical and X-ray variations}

CCD photometry was obtained during 3 nights in March 1995 with EFOSC II at 
the 2.2 m telescope on La Silla (during MPI time).
Observations were performed mostly with the $B$ filter, though a few
exposures also were made with the $V$ and $R$ filters.
Exposure times ranged between 10--600 sec.
Observational details are listed in Tab. \ref{log}.
The images were processed using the profile-fitting scheme of the 
{\sc DoPhot} reduction package (Mateo \& Schechter 1989) 
to achieve high accuracy.
The optical light curve  is characterized by a broad bright phase
and a shallow 0.3 mag deep faint phase.
In contrast, the X-ray intensity variations are much more pronounced
and show a clear on/off-like behaviour.

In order to derive an orbital period we carried out a period search using the 
analysis-of-variance method (Schwarzenberg-Czerny 1989) for both, the
optical photometry data as well as the combined \ros\ PSPC pointings 
and the phase-dispersion method (Stellingwerf 1978) for the 
\ros\ all-sky survey data.
The resulting periodograms (Fig. \ref{perio}) show a variety of maxima
which are predominantly caused by the poor sampling rather than the 
intrinsic variability. 
A more closer look at the bottom two panels (optical and \ros\ pointing
data) suggests that common periods of 123, 134 or 149 min are possible
(134 min is the most reliable period from the \ros\ pointed data alone).
Inclusion of the \ros\ all-sky survey data clearly favours the 134 min
period though we caution that due to the orbital motion of the \ros\
satellite (96 min) the signal may be affected. We note also that the
actual best-fit period of the \ros\ all-sky survey data is 132.5 min, thus
being slightly smaller than the period derived from the optical and 
pointed \ros\ data. Note that for each data set (corresponding to the three 
panels in Fig. \ref{perio}) a slightly different best-fit period is obtained
though the errors (which we consider to correspond to the width of the peak)
overlap. Overall, based on the appearance of the folded light curves
(Fig. \ref{lc}) according to a criterion of ``greatest simplicity'' we feel 
that $P_{\rm orb}$ = 134$\pm$3 min is our best estimate (shaded area in 
Fig. \ref{perio}).
 
\section{A low resolution spectrum}

\subsection{Optical identification}

The identification spectrum of the likely counterpart was obtained on 
March 26, 1995 with the ESO 2.2~m telescope at La Silla/Chile
and was exposed for 30 min. We used
the EFOSC II spectrograph equipped with a 1k$\times$1k Thomson CCD detector
with a 450 \AA/mm grism (corresponding to about 8.5 \AA/pixel)
covering the optical wavelength range from  3000--9000 \AA. 
The observation was performed  under stable photometric conditions and
accompanied by measurements of the standard star LT4816. We additionally 
applied a correction using the $B$ magnitude derived from a direct image 
taken just prior to the spectroscopic observation 
to calibrate the flux with an accuracy of $\sim 10$\% (using standard
{\sc midas} procedures). By convolving the
original spectrum with functions representing the $BV\!R$ bandpasses we arrive 
at  $B=18\fm3$, $V=18\fm5$ mag and $R=18\fm0$ with a mean error of 
$\pm$0.2 mag for \59.

\begin{figure}[t]
   \psfig{figure=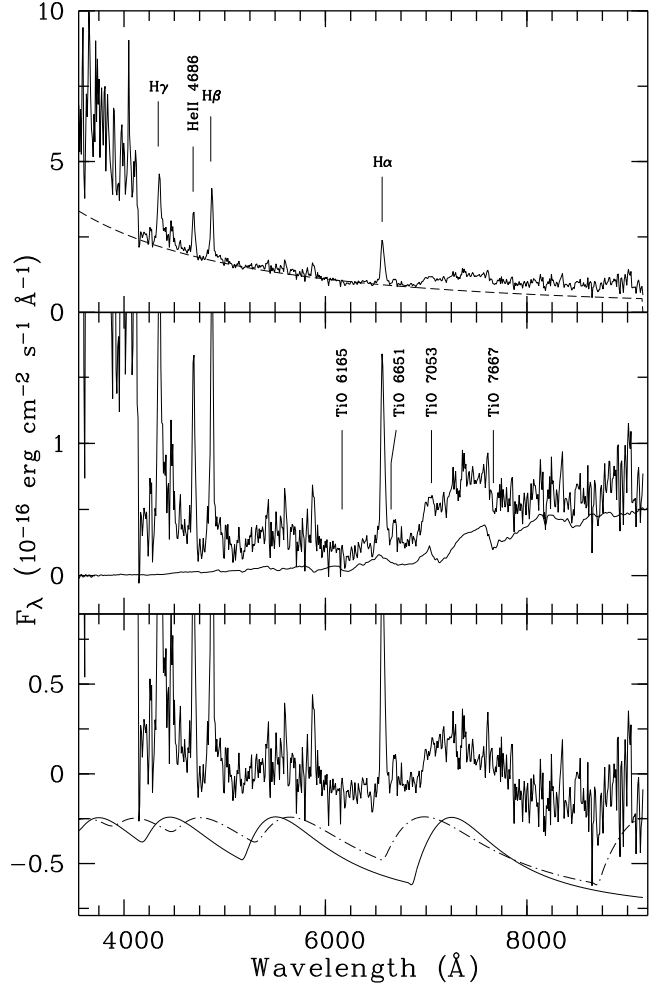,width=0.98\columnwidth,%
      bbllx=60pt,bblly=70pt,bburx=520pt,bbury=790pt,clip=}
  \caption[osp]{{\bf upper panel:}  Low resolution optical spectrum of \rxj\ 
      obtained on March 26, 1995. The dashed line represents a power-law with 
      $f_{\lambda}\propto \lambda^{-2}$ fitted in the range between 
      3800--5000 \AA. {\bf middle panel:} Residual spectrum after
      subtraction of the blue continuum shown together with the spectrum of 
      the late M-dwarf Gliese 83.1 (M~4.5) scaled to match 
      the strength of the observed TiO features. 
      {\bf lower panel:} The spectrum after the removal of the contribution of
      the  secondary star. It
      exhibits broad emission features at 5500 and 7400 \AA, tentatively 
      identified as 3rd and 4th harmonic of the cyclotron fundamental.
      The solid line represents our favoured modell with $B=52$~MG, 
      $\theta=80\degr$ and for $kT=10$~keV. The alternative interpretation
      for an $B=41$~MG field is also given (dashed line).}
  \label{ospec}
\end{figure}

The resulting  spectrum plotted in the upper panel of Fig.~\ref{ospec}
reveals strong emission lines with of the Balmer series and \ion{He}{ii} 
typical for a CV. The high-excitation line \ion{He}{ii}
$\lambda 4686$~\AA\ is quite prominent. Its strength (2/3 of the H$\beta$-flux)
and the inverted decrement of the Balmer lines 
clearly indicate a magnetic nature of the object.

\subsection{The secondary star}

We further deconvolved the optical spectrum by fitting a power-law in the blue 
range between 3800--5000 \AA. After subtracting this blue continuum 
contribution the late-type companion is revealed by the  slope of the 
continuum in the  near--IR and obvious flux depression due to the TiO 
absorption throughs which 
are best seen in the $\lambda\lambda$~7053 and 7667 \AA\ bandheads. 
The faintness of the features prevents an accurate quantitative estimate of 
the spectral type, but the low ratio $\la 0.5$ 
of the flux deficits $F_{\rm TiO}$(6165\AA) to $F_{\rm TiO}$(7667\AA) suggest
that it is later than dM~3. The expected parameters for a Roche-lobe 
filling secondary in a 134 minute system are $M_{2}= 0.18$M$_{\odot}$,
$R_{2}=0.22$R$_{\odot}$ and a spectral type between M~4.5--5.5, for which  
the ZAMS mass-radius and mass-spectral type relations 
of Patterson (1984) and Kirckpatrick \& McCarthy (1994)  were used.  

The distance of \59 can be estimated from the observed strength of the
M-dwarf using the method of Bailey (1981). In the following we take the
dM~4.5 star Gl 83.1 ($K=6\fm67$) as a template and assume that both M-stars 
have the same colour $V-K=5\fm62$ (Leggett 1992).
By comparing  the measured flux deficits in 
the $\lambda 7667$-band in Gl 83.1 and that of  \59  
(1800~10$^{16}$ and 0.2~10$^{16}$~\ergcm), we derive a $K$ magnitude 
of 16\fm5 for the secondary in \59 . If we take this value and
$S_{K}=4.43$ from the surface brightness
relationship of Beuermann \&\ Weichhold (1998) a distance of $615\pm150$ pc 
is deduced. The error includes an uncertainty of 25\% for both, the 
flux calibration  and the scaling of the TiO band deficits.

\begin{figure*}
  \psfig{figure=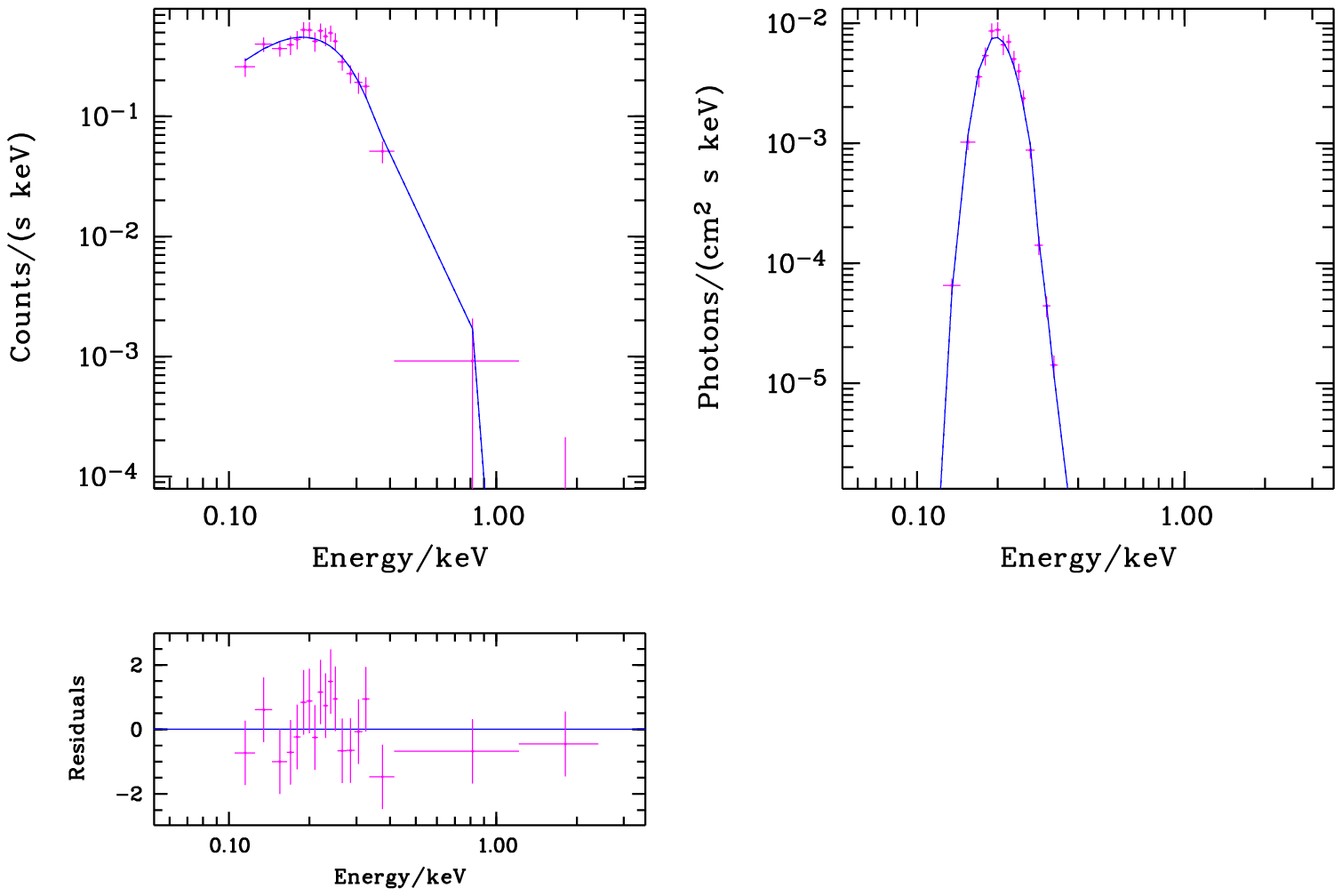,width=10.9cm,%
       bbllx=2.5cm,bblly=1.3cm,bburx=18.2cm,bbury=11.6cm,clip=}
  \vspace*{-7.cm}\hspace*{11.cm}
  \psfig{figure=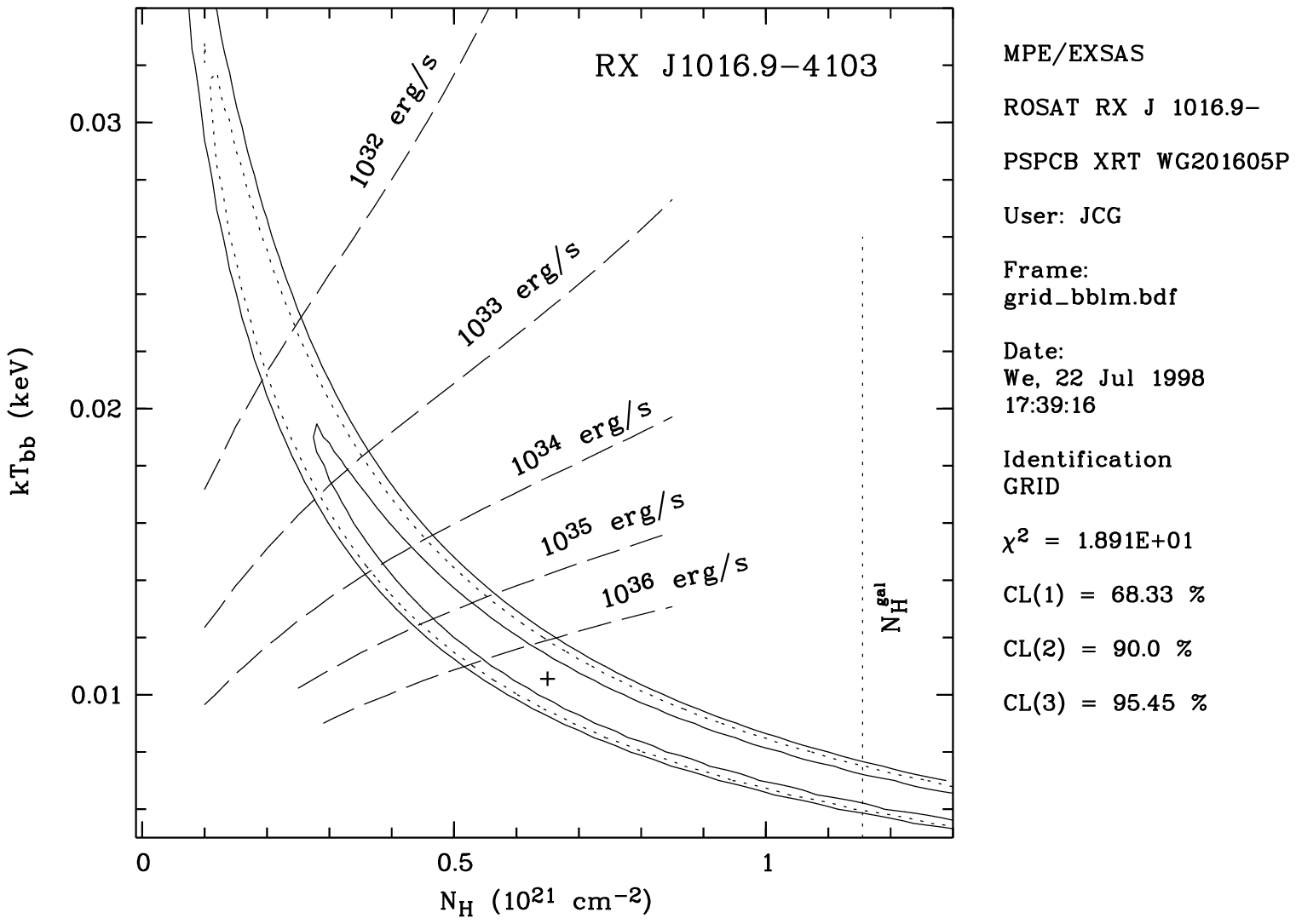,width=7.cm,%
       bbllx=1.9cm,bblly=1.1cm,bburx=13.5cm,bbury=12.1cm,clip=}
  \vspace*{0.2cm}
  \caption[xsp]{{\bf Left:} Phase-averaged X-ray spectrum of the two merged
      PSPC observations of \59\ unfolded with the sum of a blackbody and a
      thermal bremsstrahlung spectrum with a temperature fixed at 20 keV 
      (see text for details).  The lower
      left panel shows the residua of the fit in units of $\sigma$.
      {\bf Right:} The 68\%, 90\% and 95\% significance contours
          of the blackbody model fit to the merged PSPC spectrum of \59\
         in the $N_{\rm H}-kT$ plane. Overplotted are the lines of constant 
       luminosity in units of (D /100 pc)$^2$ (dashed), as well as the value 
       of the total galactic absorbing column (vertical dotted line) 
       according to Dickey \& Lockman (1990).   }
  \label{xspec}
\end{figure*}

Detailed phase-resolved spectroscopic studies of selected systems,
e.g. AM Her (Davey \& Smith 1996), QQ Vul (Schwope \etal\ 1998) and
RX J0203.8+2959 (Schwarz \etal\ 1998)
have demonstrated that absorption features from the illuminated side 
of the secondaries in magnetic CVs can be heavily suppressed due 
to the strong X/UV irradiation from the accretion region.
We therefore caution that the distance derived from our single spectrum with 
unknown viewing geometry of the binary system can serve as an upper limit only.

\subsection{A magnetic field estimate}

After subtraction of a suitably scaled spectrum of Gl 83.1 two broad
emission humps at $\lambda\lambda$ 5500 and 7400 \AA\ clearly stand out
in the red part of the spectrum (Fig.~\ref{ospec}, lower panel), which 
we tentatively identify as cyclotron lines from the hot optically 
thin post-shock region.

An unambiguous estimate of the magnetic field strength is difficult to
achieve with only two distinct humps observed. Their separation 
favours an interpretation as the 3rd and 4th harmonic of the cyclotron
fundamental, in which case the field strength is 52$\pm2$~MG slightly 
depending on the plasma temperature and the yet unknown field orientation.
An alternative interpretation as  the 4th and 5th harmonics, in which
case $B$ is $\sim41$~MG, cannot not completely excluded but results in a
distinct mismatch of the expected humps with the shape of the continuum
between 6000--7000 \AA. Normalized cyclotron absorption coefficients for 
both cases assuming  $\theta = 80\degr$ and $kT = 10$~keV
shown in Fig.~\ref{ospec} illustrate this ambiguity.

\section{The X-ray spectrum}

The two pointed PSPC observations have resulted in the detection of nearly
800 photons thus enabling  spectral investigation.
For the spectral analysis the source photons of the two PSPC pointings
were merged and then extracted with a radius of 1\farcm5. The background was 
chosen from concentric circles around the source region with a radius of 
3\amin. Other nearby sources were cut out, and the background area 
normalized to the source extraction area before the background subtraction.   
Standard corrections were applied using the dedicated EXSAS software package
(Zimmermann \etal\ 1994).
As a first step, we have considered the total spectrum as collected
during each of the two observations (after merging). 
Adopting a blackbody plus a thermal bremsstrahlung model (the temperature
of the latter fixed to 20 keV) we find that there 
is no need at all for the inclusion of the bremsstrahlung component,
that is the spectral fit does not require any spectral component for hard
emission (Fig. \ref{xspec}). The resulting best-fit
temperature of the blackbody component is  10$\pm$7 eV and 
the absorbing column $N_{\rm H}$=6.4$\times$10$^{20}$ cm$^{-2}$, e.g.
about 50\% of the total galactic absorbing column in this direction 
(Dickey \& Lockman 1990). However, this gives an unrealistic high luminosity
for a cataclysmic variable due to the very low temperature.
If we fix the temperature at the canonical value of 20 eV, the best-fit
absorbing column is only $N_{\rm H}$=2.4$\times$10$^{20}$ cm$^{-2}$.
The difference in reduced $\chi^2$ is only marginal for all temperatures 
in the 10--25 eV range. 

As a second step we have investigated the emission during the off-state
phase intervals which has been observed for a total of 7340 sec.
Selecting time intervals where the 60 sec averaged X-ray intensity is
less than 0.05 cts/s, we find that the residual emission is consistent
with being background radiation. The mean count rate during the X-ray bright
phase (at $>$0.05 cts/s) is 0.38 cts/s in the PSPC.

Considering only the X-ray bright phase, the unabsorbed fluxes of the two 
model components in the \ros\ band (0.1--2.4 keV) are    
$F_{\rm bbdy}$ = 5.1$\times10^{-11}$ \ergcm\ (using kT$_{\rm bbdy}$=20 eV
 for comparison purposes) and
$F_{\rm thbr}$ $<$9.8$\times$10$^{-13}$ \ergcm, giving a small flux ratio 
of $F_{\rm thbr}$/$F_{\rm bbdy} < 0.02$. The mean bolometric luminosity 
 during the X-ray bright phase is (again with kT$_{\rm bbdy}$=20 eV)
$L_{\rm X}$ = 2.4$\times$10$^{32}$ (D / 100 pc)$^2$ erg/s.

\section{Discussion and Conclusion}

The measured period of $P=134$ min places \59\ in the 2--3 hr CV period gap
which is thought to be due to the transition from orbital angular momentum
loss by magnetic braking (P$>$ 3 hr) to gravitational radiation (P$<$ 2 hr)
(King 1988). Over the last years, more polars have been found with periods
in the 2--3 hr range thus re-initiating the debate on the existence and
significance of the period gap for magnetic systems (Wickramasinghe \& Wu 1994,
Wheatley 1995).

The shape of the X-ray light curve strongly argues for a self-eclipsing polar.
The length of the X-ray bright phase of $\la 0.5$  phase units suggests
(within our limited accuracy) an one pole accreting geometry such that
the accretion region passes behind the limb of the white dwarf for half
the orbital period. The lack of eclipses implies $i < 78\degr$.

Our tentative estimate of the magnetic field strength is 52$\pm2$~MG
and the strong soft X-ray excess are in good agreement with the
correlation found between these quantities (Beuermann 1998).

The strongly modulated soft X-ray emission, the seemingly synchronous rotation,
the strength and relative intensities of 
the Balmer and \ion{He}{ii} emission lines as well as the cyclotron humps in 
the optical spectrum clearly suggest the polar nature of \59\ though
no polarimetric measurements have been made. More detailed optical
photometry and phase-resolved spectroscopy as well as polarimetry are 
needed to determine the system parameters of this new polar.

\begin{acknowledgements}
We thank R. Egger for help during the ESO observing run and A.D. Schwope for 
generously providing help and software for the cyclotron spectroscopy.
JG and RS are supported by the Deut\-sches Zentrum f\"ur Luft- und Raumfahrt 
(DLR) GmbH under contract No. FKZ 50 QQ 9602\,3 and 50 OR 9206\,8. 
The \ros\ project is supported by the German Bundes\-mini\-ste\-rium f\"ur 
Bildung, Wissenschaft, For\-schung und Technologie (BMBF/DLR) and the 
Max-Planck-Society.
\end{acknowledgements}


\begin{thebibliography}{}

\bibitem[]{bai81} Bailey J., 1981, MNRAS 197, 31

\bibitem[]{bb95} Beuermann K., Burwitz V., 1995, ASP Conf. Ser. 85, 99

\bibitem{b96} Beuermann K., 1998, in Perspectives of High Energy
   Astronomy \& Astrophysics, Proc. of Int. Coll. to commemorate the Golden 
   Jubilee of TIFR, Tata Inst. of Fund. Research, India, Aug. 1996, p. 100

\bibitem[]{bw98} Beuermann K., Weichhold M., 1998, A\&A (in press)

\bibitem[]{brs96} Burwitz V., Reinsch K., Schwope A.D., Beuermann K., 
Thomas H.-C., Greiner J., 1996, A\&A 305, 507

\bibitem{ds94} Davey S.C., Smith R.C., 1996, MNRAS 280, 481

\bibitem{dl90} Dickey J.M., Lockman F.J., 1990, ARAA 28, 215

\bibitem[]{g95} Greiner J., Remillard R., Motch C.,  1995, in Cataclysmic 
Variables, eds. A. Bianchini,  M. Della Valle, M. Orio, ASSL 205, p. 161

\bibitem[]{g96} Greiner J., 1996, in Supersoft X-ray Sources,
ed. J. Greiner, Lecture Notes in Physics 472, Springer, p. 285

\bibitem[]{grm98} Greiner J., Remillard R., Motch C.,  1998, A\&A 336, 200

\bibitem[]{gsw98} Greiner J., Schwarz R., Wenzel W., 1998, MNRAS 296, 437

\bibitem[]{king88} King A.R., 1988, QJRAS 29, 1

\bibitem[]{kirk94} Kirkpatrick J.D., McCarthy D.W., 1994, AJ 107, 333 

\bibitem{kp82} Kujpers J., Pringle J.E., 1982, A\&A 114, L4

\bibitem[]{} Leggett S.K., 1992, ApJS 83, 351 

\bibitem[]{ms89} Mateo M., Schechter P., 1989, 1st ESO/ST-ECF Data Analysis 
              Workshop, eds. Grosbol P.J., Murtagh F. \& Warmels R.H., p. 69

\bibitem[]{patt84} Patterson J., 1984, ApJS 54, 443

\bibitem[]{ssbb98} Schwarz R., Schwope A.D., Beuermann K., \etal , 1998,
   A\&A 338 (in press)

\bibitem[]{s89} Schwarzenberg-Czerny A., 1989, MNRAS 241, 153

\bibitem[]{sbcms98} Schwope A.D., Beuermann K.,  Catal\'{a}n M.S., Metzner A.,
  Steeghs D., 1998, MNRAS (subm.)

\bibitem{s78} Stellingwerf R.F., 1978, ApJ 224, 953

\bibitem[]{w95} Warner B., 1995, Cataclysmic Variable Stars, Cambridge 
              Univ. Press, Cambridge

\bibitem[]{wh95} Wheatley P.J., 1995, MNRAS 274, L51

\bibitem[]{ww94} Wickramasinghe D.T., Wu K., 1994, ApSS 211, 61

\bibitem[]{wb96} Woelk U., Beuermann K., 1996, A\&A 306, 232

\bibitem[]{zbb94} Zimmermann H.U., Becker W., Belloni T., \etal,
1994, MPE report 257

\end{thebibliography}
\end{document}